\documentclass[aps,prb,amsmath,twocolumn,amssymb,titlepage]{revtex4-1}
\usepackage{graphicx}
\usepackage{wrapfig}
\usepackage{nicefrac}
\usepackage{amsfonts}
\usepackage{verbatim}   
\usepackage{color}      
\usepackage[usenames,dvipsnames]{xcolor}
\usepackage[caption=false]{subfig}
\usepackage{amssymb}
\usepackage{amsmath} 

\usepackage{multirow} 
\usepackage{tabularx} 
\usepackage{array}

\usepackage{units}
\usepackage{comment}
\usepackage{tensor} 
\usepackage{braket}
\usepackage{bm}
\usepackage{natbib}
\usepackage{placeins}
\bibpunct{[}{]}{,}{n}{}{}
\usepackage[hidelinks]{hyperref} 
\hypersetup{colorlinks=true,citecolor=Red,linkcolor=Blue,urlcolor=Blue}

\usepackage[export]{adjustbox}

\renewcommand{\Im}{\operatorname{{\mathrm Im}}}

\renewcommand{\>}{\rangle}
\newcommand{\<}{\langle}

\newcommand{\w}{\omega}

\newcommand{\tcite}[1]{\textcolor{red}{ \bf{cite} }}

\begin{document}

\title{Two-Magnon Bound States in the Kitaev Model in a $[111]$-Field}
\author{Subhasree Pradhan}
\email[E-mail:]{pradhan.61@osu.edu}
\author{Niravkumar D. Patel}
\email[E-mail:]{patel.3537@osu.edu}
\author{Nandini Trivedi}
\email[E-mail:]{trivedi.15@osu.edu}
\affiliation{Department of Physics, The Ohio State University, Columbus, OH-43210, USA}

\begin{abstract}
It is now well established that the Kitaev honeycomb model in a magnetic field along the $[111]$-direction harbors an intermediate gapless quantum spin liquid (QSL) phase sandwiched between a gapped non-abelian QSL at low fields $H< H_{c1}$ and a partially polarized phase at high fields $H> H_{c2}$. Here, we analyze the low field and high field phases and phase transitions in terms of single- and two-magnon excitations using exact diagonalization (ED) and density matrix renormalization group (DMRG) methods. We find that the energy to create a bound state of two-magnons $\Delta_p$ becomes lower than the energy to create a single spin flip $\Delta_s$ near $H_{c2}$. In the entire Kitaev spin liquid $\Delta_p<\Delta_s$ and both gaps vanish at $H_{c2}$. We make testable predictions for magnon pairing that could be observable in Raman scattering measurements on Kitaev QSL candidate materials.
\end{abstract}

\date{\today}
\maketitle

\noindent {\it Introduction:} Quantum spin liquids (QSL) have generated significant excitement because of their potential applications for topological quantum computation. In QSLs, magnetic order is suppressed by quantum fluctuations hence they cannot be described within the traditional Landau theory of symmetry-breaking which is based on the existence of a local order parameter. Instead, the concept of topology plays a central role in the study of QSLs. QSLs are characterized by long-range entanglement, multiple degeneracy of the ground-state, fractionalized quasi-particles and the existence of topological order~\cite{QSL1, QSL2, Balents, Mila, 2dQSL, You}. 
 
\begin{figure*}[t]
\centering
\includegraphics[trim={0cm 0cm 0cm 0cm},width=0.9\linewidth]{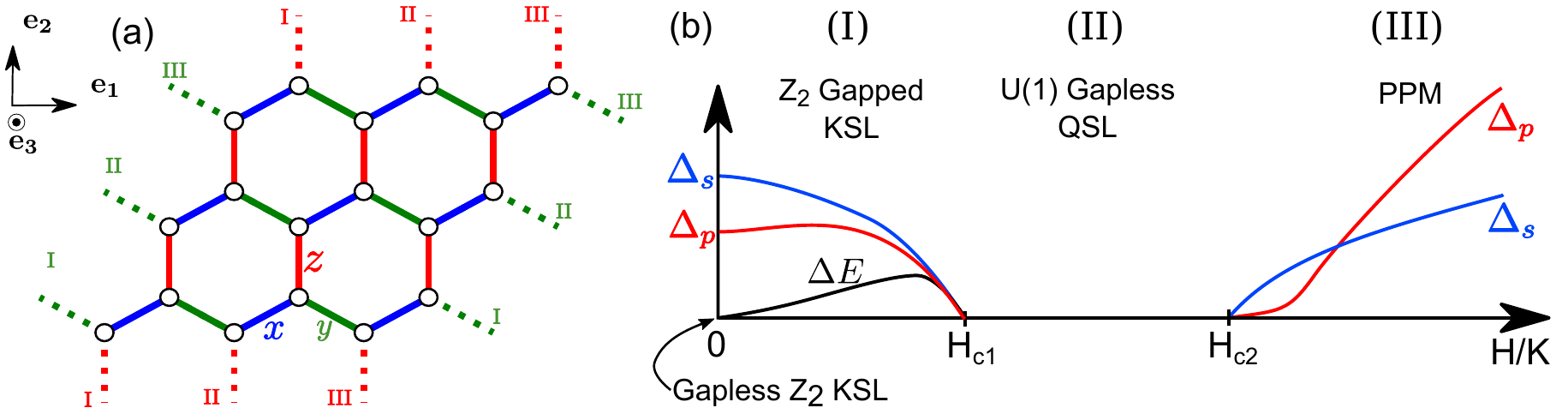}
\subfloat{\label{fig:1a}}
\subfloat{\label{fig:1b}}
\caption{(a) The Kitaev honeycomb model with Ising exchange coupling between Pauli spin operators along $x$ (blue), along $y$ (green) and along $z$ (red) bonds. Vectors $\hat e_1$, $\hat e_2$ and $\hat e_3$ represent the cartesian coordinates on the honeycomb lattice. (b) Schematic phase diagram based on references~\cite{Nirav, Trebst} that show the evolution of the gapless $Z_2$ Kitaev spin liquid (KSL) to a $Z_2$ gapped KSL (I) as a function of a magnetic field along $[111]$ or $e_3$ perpendicular to the honeycomb plane. The one-particle gap $\Delta_s$ is the energy cost for creating a spin-flip or a magnon excitation with a change in the spin quantum number $\Delta S = 1$ . The two-particle gap $\Delta_p$, main result of our paper, is the energy cost of creating two spin-flips or a two-magnon bound-state. A phase transition is signaled by closing of $\Delta_s$ and $\Delta_p$ gaps with increasing field, leading to an intermediate $U(1)$ gapless QSL (II). For $H > H_{c2}$, the gaps $\Delta_s$ and $\Delta_p$ become finite again in the partially polarized magnetic (PPM) phase (III) and increase linearly at high fields $H \gg H_{c2}$. In the region $H < H_{c1}$, $\Delta_p$ is always smaller than $\Delta_s$. In the PPM phase, there is a cross over between the energy scales $\Delta_s$ and $\Delta_p$ that occurs near $H_{c2}$. The schematic plots for $\Delta_s$ and $\Delta_p$ are based on calculations presented in fig.~\ref{fig:2c}. The analysis of the lowest energy gap $\Delta E$, magnetization $M$ and the peaks in magnetic susceptibility $\chi$ as a function of $H$ give $H_{c1}=0.21$ and $H_{c2}=0.34$ in Fig. S1 in Supplementary Material.}
\label{fig:1}
\end{figure*}

The Kitaev spin-$1/2$ model on the honeycomb lattice~\cite{Kitaev} is the paradigmatic example for a QSL because of its unique combination of exact solvability hosting a variety of gapped and gapless QSL 
phases~\cite{Knolle_thesis,Nirav,David,Sheng,hickey2018gapless,YuanMing}
and for having experimental relevance~\cite{Khaliullin,Trebst,Winter2017,chun2015,Yogesh,Banerjee1055,Banerjee2018, Kasahara2018,Martin1}. 
It is described by 
\begin{equation} \label{eq:Kitaev}
H_{K} = \sum_{\gamma = {x, y,z}}  K^{\gamma} \sum_{\< i j \>_{\gamma}} S_i^\gamma S_j^\gamma, 
\end{equation}
where we take the interaction parameter $K$ to be antiferromagnetic ($K^{\gamma}>0$). The pairwise nearest neighbor Ising spin interactions are bond $(\gamma = x,y,z)$-dependent between sites $i$ and $j$ (Fig.~\ref{fig:1a}).
The isotropic AFM Kitaev model ($K^x = K^y = K^z = 1 \text{eV}$) has a topologically non-trivial gapless QSL ground-state. Following Kitaev's original solution, each spin-$1/2$ can be split into four Majorana fermions: three are associated with the bonds and one with the original site. The bond Majoranas can be recombined to form a static $Z_2$ gauge field, leaving a single free Majorana fermion moving in a background of $Z_2$-gauge fields. The Majorana spectrum is gapless with Dirac points located at the $K/K'$ points of the Brillouin zone, yielding a gapless $Z_2$ Kitaev spin liquid (KSL)~\cite{Kitaev,Knolle_thesis}. 

In this Letter our main goal is to obtain the effect of a magnetic field on the magnetic excitation spectrum. As shown schematically in Fig.~\ref{fig:1b}, we have previously discovered two transitions between a gapped KSL and a gapless $U(1)$ QSL at $H_{c1}$ and a second phase transition between the gapless $U(1)$ QSL and a partially polarized magnetic phase at $H_{c2}$~\cite{David,Nirav}. We choose to use the hard core boson (HCB) representation~\cite{suppl} to describe the $S = 1/2$ operators in order to describe the gap closing at the critical fields in the familiar language of multi-magnon excitations. Our main contribution is the calculation of the dynamical one- and two-particle spectra as a function of magnetic field from which we extract the gap scales as shown schematically in Fig.~\ref{fig:1b}.

\bigskip

\noindent {\it Kitaev model in a magnetic field along [111]:}
The isotropic AFM Kitaev model with an external magnetic field applied in the $[111]$-direction is defined by adding $-\mathbf{H} \cdot  \sum_{i \gamma} \mathbf{S_i}^{\gamma}$ to the Kitaev Hamiltonian in Eq. \ref{eq:Kitaev}, where $\mathbf{H} = H(\hat{e}^x+\hat{e}^y+\hat{e}^z)$ is perpendicular to the 2D honeycomb plane with equal projections along the bond directions $\{\hat{e}^x,\hat{e}^y,\hat{e}^z\}$.
We use density matrix renormalization group (DMRG)
~\cite{WhiteDMRG1,WhiteDMRG2,WhiteWFT,Springerbook1,
alvarez0209,alvaez3,Schollwock1,Schollwock2} 
to directly simulate the interacting spin model and exact diagonalization (ED) to evaluate the spectrum of the HCB model. The dynamical spectra are obtained using Lanczos on small clusters~\cite{Springerbook1,Lanczos}. Overall, the combination of the spin and hard-core boson representations provides useful insights.\\

\noindent {\it One- and Two-spin Dynamical Spectra:}
We calculate the one-particle (magnon) and two-particle (two-magnon) dynamical spectra as a function of the magnetic field $H$. Inelastic neutron scattering (INS) spectroscopy gives information about the magnon dispersion. Here we make predictions for two-magnon spectroscopy that can be probed by Raman spectroscopy to see the effects of magnon-magnon bound states~\cite{TomDevereaux1}. To this end, we calculate the magnon density of states $S(\w)$ and the magnon pair density of states $P(\omega)$ defined by 
%
\begin{equation}
    \begin{split}\label{eq:one-two particle}
        S(\omega) &= \frac{-1}{N \pi} \Im 
        \Bigg[ \sum_{\substack{m \ne 0 , i \\ \alpha = +, -, z}} 
        \frac{ |\<0| S^{\alpha}_i|m\>|^2}{\omega +E_{0}-E_m + i\eta} 
        \Bigg], \\
        P^{\gamma}(\omega) &= \frac{-1}{N \pi} \Im 
        \Bigg[ \sum_{\substack{m \ne 0 , i \\ \alpha = +, -, z}} 
        \frac{ |\<0| S^{\alpha}_i  S^{\alpha}_{i+{\gamma}}|m\>|^2}{\omega + E_{0}- E_m + i\eta} 
        \Bigg],
    \end{split}
\end{equation}
%
\noindent
where, $N$ is the total number of sites, $m$ labels the eigenstate with energy $E_m$ (with $m = 0$ being the ground-state), $\gamma = \{x,y,z\}$ is the bond direction with respect to site $i$, $\omega$ is the energy transferred to the system. We use fixed $\Delta \w = 0.01$ with $\eta=0.02$ as an artificial broadening parameter for all presented calculations.
We consider all possible spin-flip processes built in through $\alpha = {\{+, -, z}\}$. The one-magnon creation operator $S_i^+$ scatters a boson from an initial state $\ket 0$ and to a final state $\ket m$.
The Raman scattering involves the creation or destruction of a pair of magnons denoted by a pair of spin operators $ S^{\alpha}_i  S^{\alpha}_{i+{\gamma}}$ along different bonds $\gamma$.

\begin{figure*}
\centering
\begin{minipage}{.33\textwidth}
\includegraphics[trim={0.0cm 0cm 0 0cm},clip,width=1\linewidth]{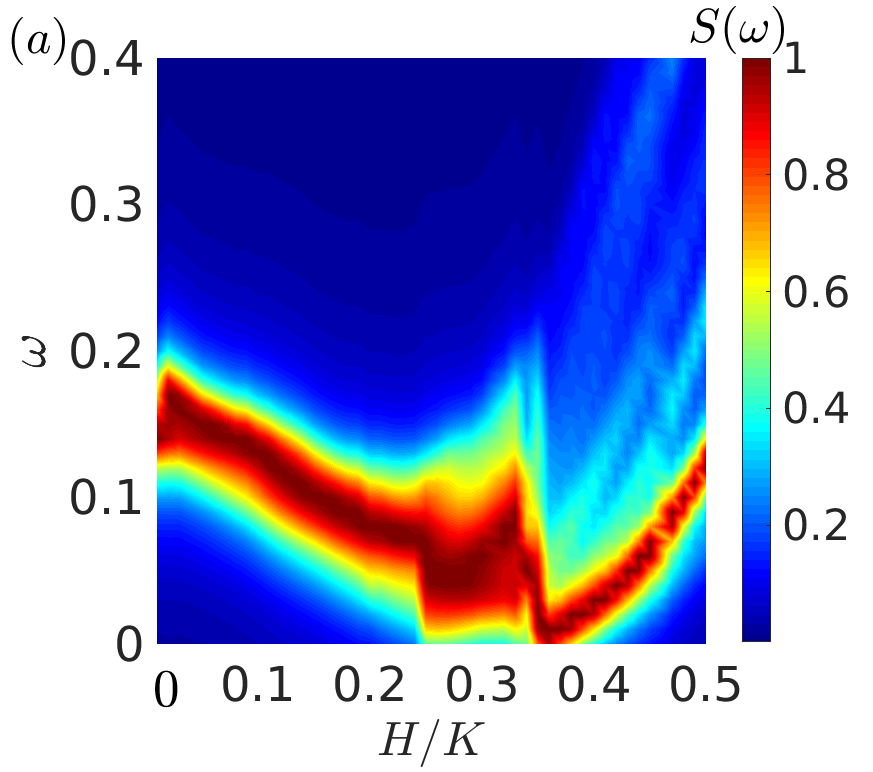}
\end{minipage}%
\begin{minipage}{.33\textwidth}
\includegraphics[trim={0.0cm 0cm 0 0},clip,width=1\linewidth]{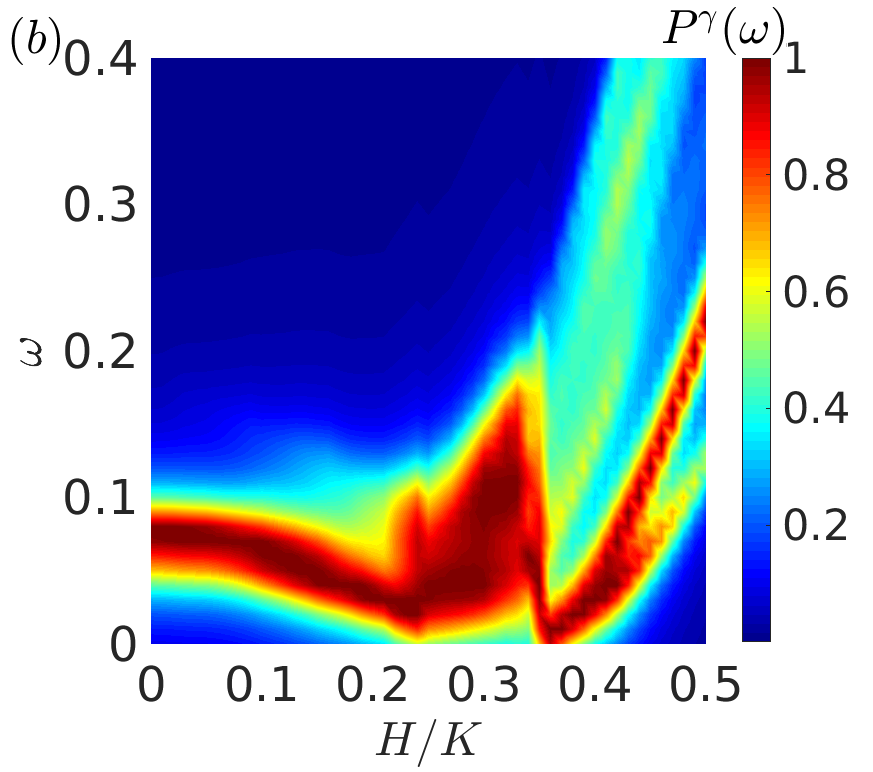}
\end{minipage}
\begin{minipage}{.33\textwidth}
\includegraphics[trim={0.0cm 0cm 0 0},clip,width=1\linewidth]{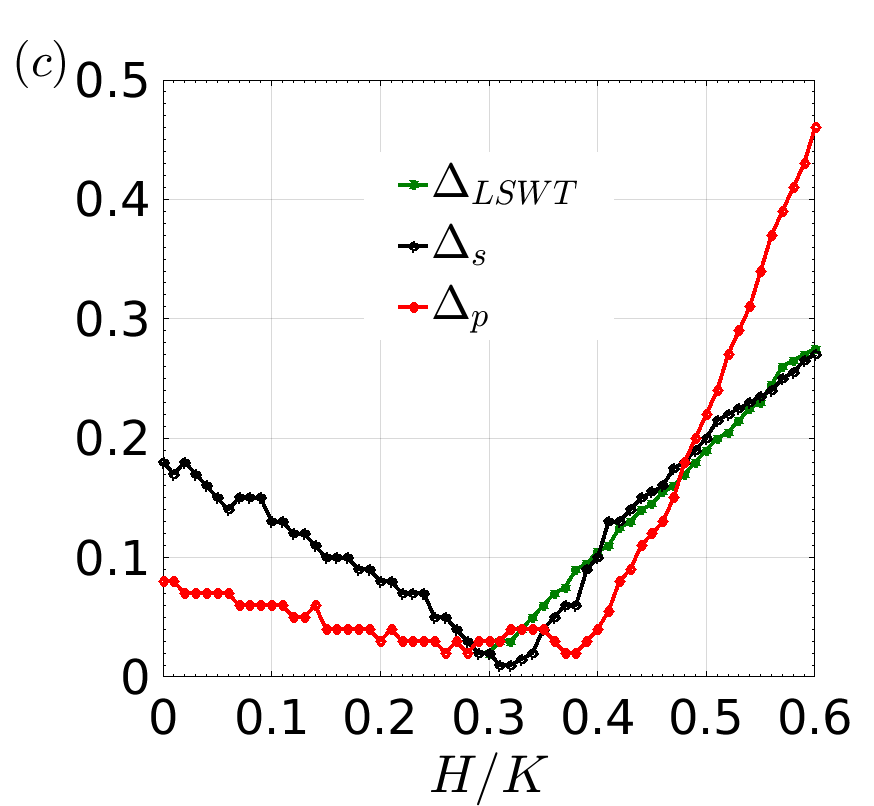}
\end{minipage}
	\subfloat{\label{fig:2a}}
    \subfloat{\label{fig:2b}}
    \subfloat{\label{fig:2c}}
\vspace{-0.76cm}
\caption{Contour plot for the normalized (a) one-particle magnon density of states $S(\omega)$ and (b) two-particle pair density of states $P^{\gamma}(\omega)$. (c) Single-particle and two-particle gap as extracted from (a) and (b) and the green line shows gap extracted from LSWT calculation which agrees quite well at the high-field limit. 
}
\label{fig:2}
\end{figure*}

\begin{figure*}[t]
\centering
\begin{minipage}{.33\textwidth}
\includegraphics[trim={0cm 0cm 0 0},clip,width=1\linewidth]{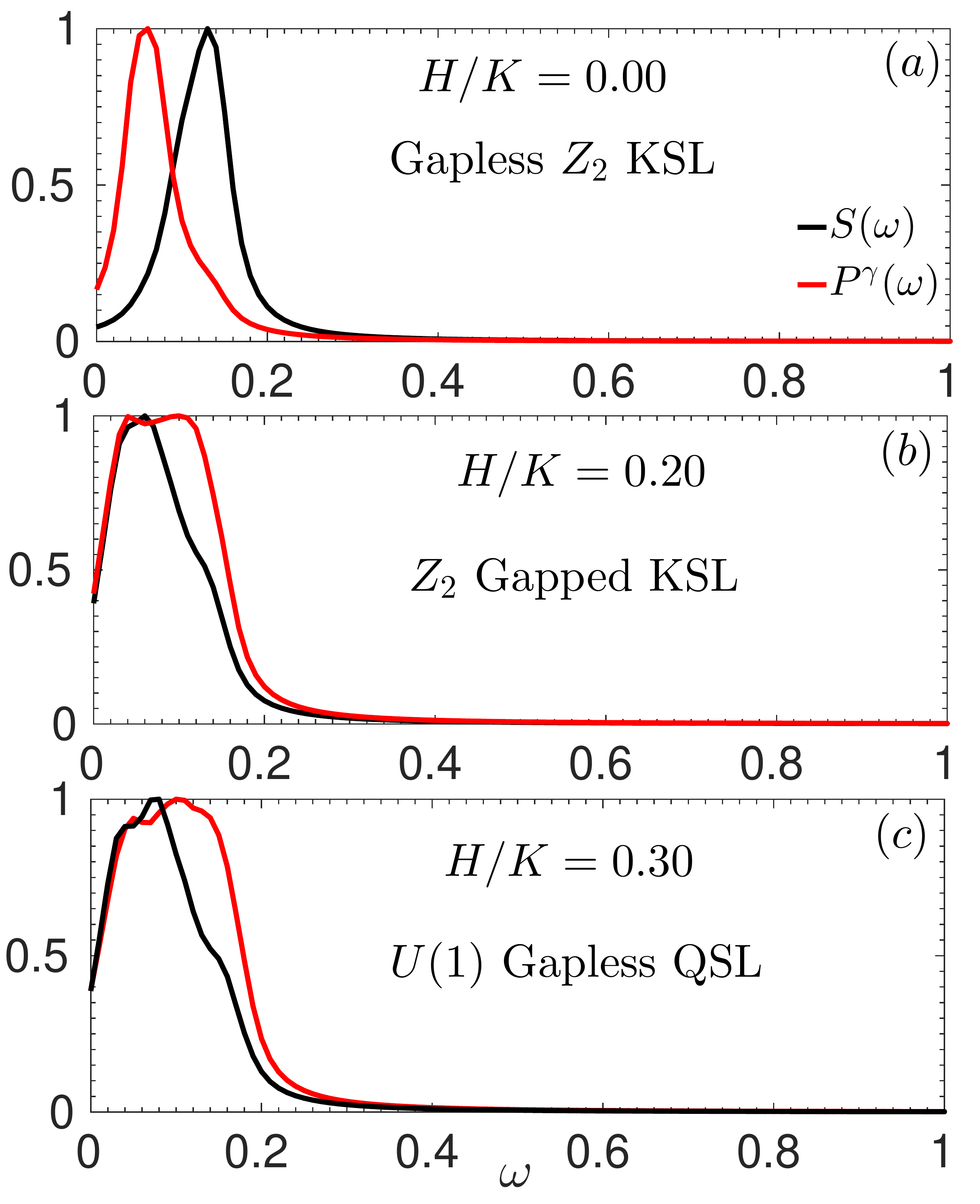}
\end{minipage}%
\begin{minipage}{.33\textwidth}
\includegraphics[trim={0cm 0cm 0 0},clip,width=1\linewidth]{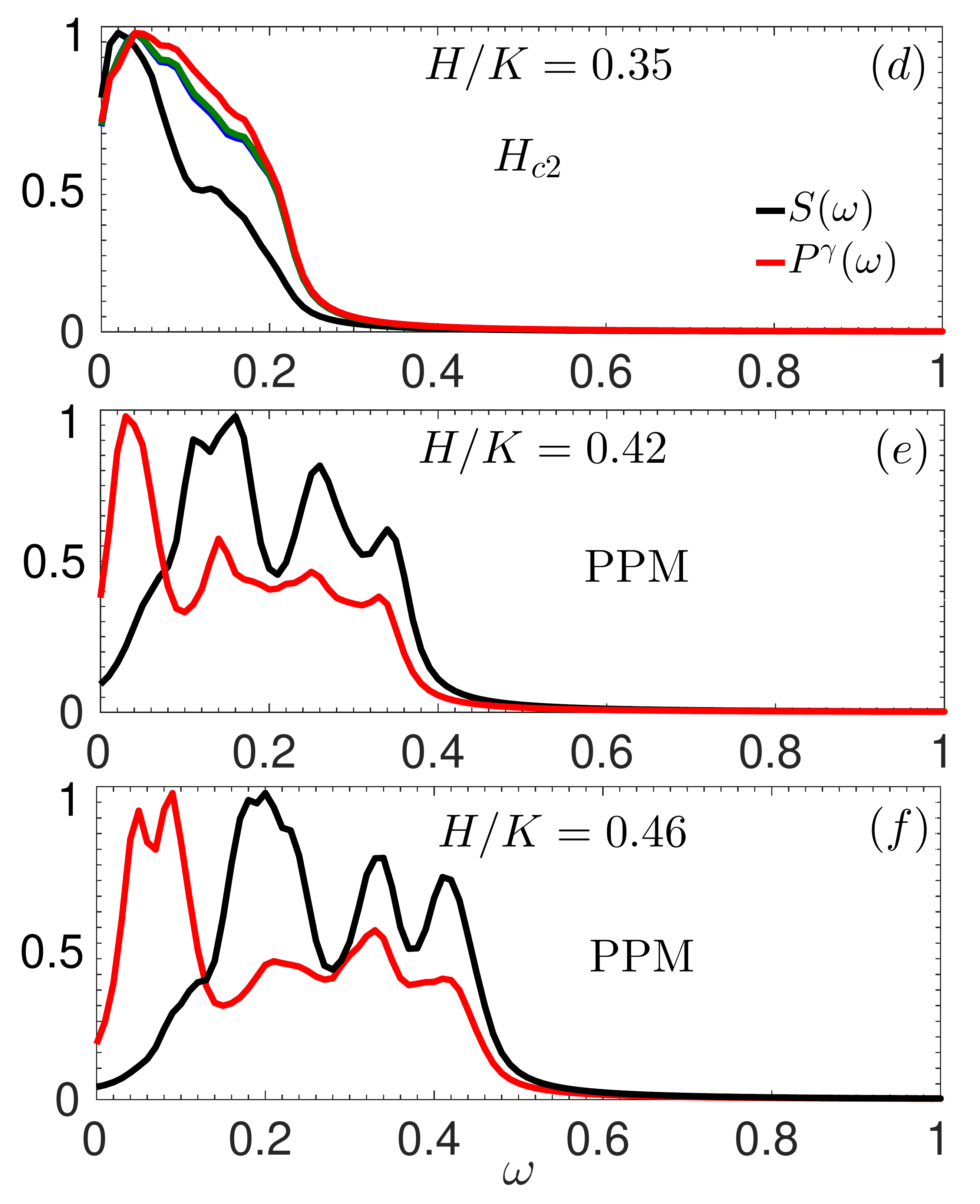}
\end{minipage}
\begin{minipage}{.33\textwidth}
\includegraphics[trim={0cm 0cm 0 0},clip,width=1\linewidth]{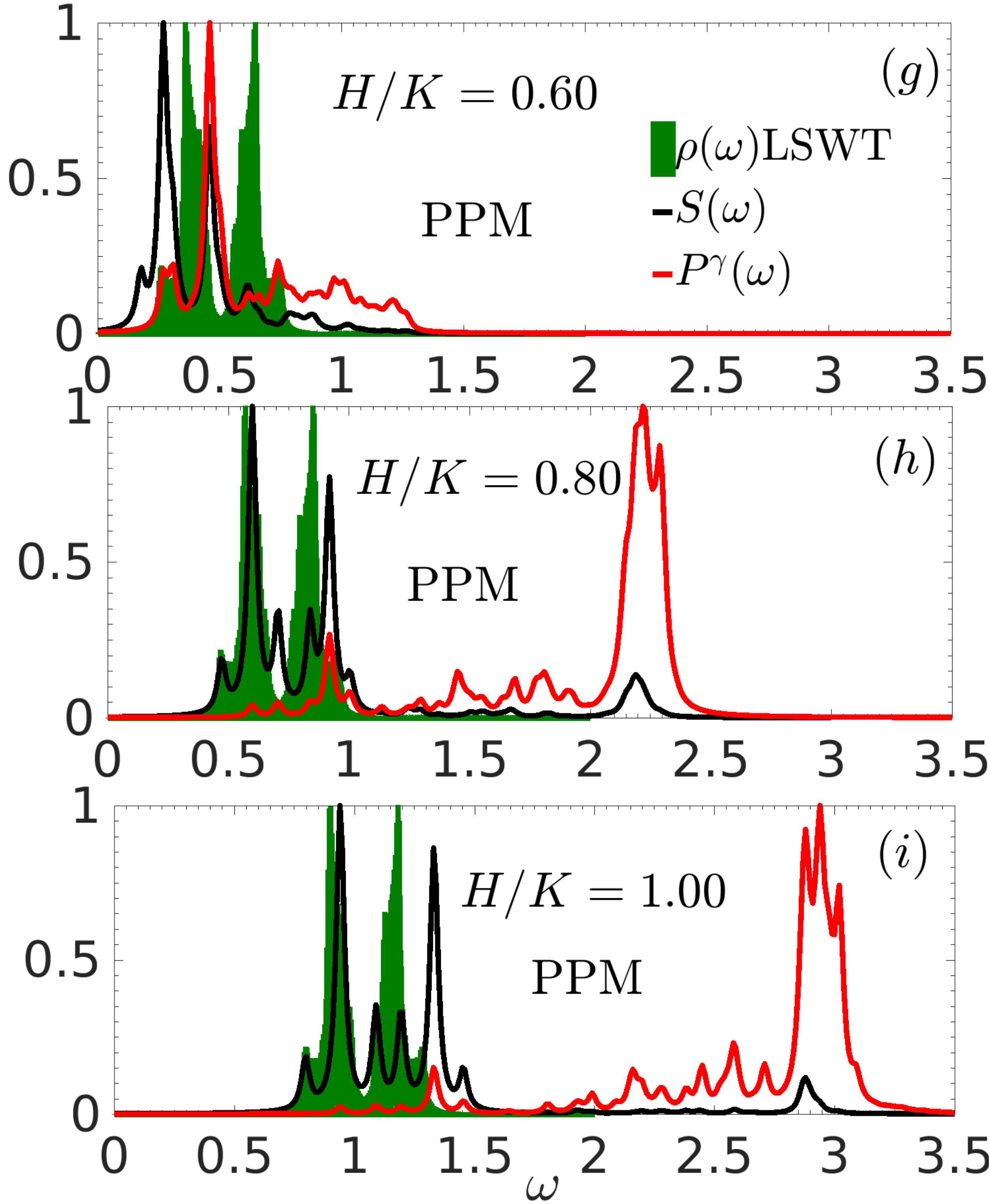}
\end{minipage}
	\subfloat{\label{fig:3a}}
    \subfloat{\label{fig:3b}}
    \subfloat{\label{fig:3c}}
    \subfloat{\label{fig:3d}}
    \subfloat{\label{fig:3e}}
    \subfloat{\label{fig:3f}}
    \subfloat{\label{fig:3g}}
    \subfloat{\label{fig:3h}}
    \subfloat{\label{fig:3i}}
\vspace{-0.6cm}
\caption{
The normalized one-magnon $S(\omega)$ and two-magnon $P^{\gamma}(\omega)$ dynamical spectra are shown for certain cuts in the contour plot of the intensity. Left panel (a)-(c) shows one and two-particle excitations from zero-field Kitaev limit towards gapless phase. The middle panel (d)-(f) $H_{c2}$ to partially polarized phase and the right panel (g)-(i) gives the magnon intensity for the high-field polarized phase. DOS calculated from LSWT is shown for high-field phases where the agreement is good.}
\label{fig:3}
\end{figure*}

Fig.~\ref{fig:2} shows density of states (DOS) plots of one-particle DOS $S(\omega)$, two-particle DOS $P(\omega)$ and the corresponding gaps $\Delta_{s/p}$ for all $H$. We also show $S(\omega)$ and $P(\omega)$ for a few $H$ within each phase of the phase diagram in Fig.~\ref{fig:3}. We also compare one-particle spectra obtained by exact methods with the corresponding spectra and gap $\Delta_{LSWT}$ obtained from linear spin-wave theory (LSWT)~\cite{suppl}. 

\noindent {\underline{$H<H_{c1}$}}:
For $H = 0$ in the gapless $Z_2$ KSL, the two particle gap is less than the energy to create a single spin flip (SF), $\Delta_p < \Delta_s$, (Figs.~\ref{fig:2c} and \ref{fig:3a}). With increasing $H$, the system enters the $Z_2$ gapped KSL phase in which $\Delta_p$ continues to remain below $\Delta_s$ and both gaps vanish at $H_{c1}$ (see Figs.~\ref{fig:3b} and \ref{fig:2c}).

\noindent {\underline{$H_{c1}\le H \le H_{c2}$}}: A broad high energy continuum in the one and two-particle excitation is found for the intermediate gapless QSL (Figs.~\ref{fig:2}a-b and \ref{fig:3}c-d), that extends upto $\w \simeq 0.15$ for $S(\w)$ and $\w \simeq 0.2$ for $P(\w)$. We attribute this characteristic high energy continuum to the formation of a Fermi surface of fractionalized charge neutral spin-1/2 spinons that are coupled to U(1) gauge field fluctuations~\cite{Nirav}. We show here that in addition to gapless spin-flip excitations, two spin-flip excitations are also gapless, $\Delta_p = 0$.

\noindent {\underline{$H>H_{c2}$}}: 
The continuum splits into multiple independent modes that shift to higher energies with increasing $H$ (Figs.~\ref{fig:2} and \ref{fig:3}e-i). 
For sufficiently large magnetic field ($H > H_{c2}$), the ground state gets polarized along the direction of $\mathbf{H}$, leading to partially polarized magnetic (PPM) phase that is precursor to a trivially 
polarized product state. This transition is signaled by opening of a spin-gap $\Delta_s$ that increases linearly with field at large $H$. At high fields the two-magnon gap $\Delta_p=2 \Delta_s$, as expected, however, close to the transition, we find a crossover between gaps with $\Delta_p$ dipping below $\Delta_s$ (see Fig.~\ref{fig:2c} and~\ref{fig:3}e-f).
Further away from the transition, Fig.~\ref{fig:3}g-i shows that the two-particle spectral weight shifts to higher energies compared to the one-particle indicating a crossover of the gap scales.

Fig.~\ref{fig:2c} shows one particle $\Delta_s$ and two-particle $\Delta_p$ gaps as a function of $H$, extracted from the exact calculations. 
Fig.~\ref{fig:2c} shows remarkable agreement between $\Delta_{LSWT}$ and $\Delta_{s}$ for sufficiently large fields. Additionally, we demonstrate that there is a crossover in $\Delta_s$ and $\Delta_p$ at $H\simeq 0.5$ where $\Delta_p < \Delta_s$. This is one of the key results of our work. We demonstrate that two-particle excitations play a main role near the $H_{c2}$ phase transition. 
In fact, Figures~\ref{fig:3}a-b show that low energy physics is dominated by (both) one and two magnon excitations. 
For lower fields, $H_{c2}<H<0.4$, the LSWT results significantly deviate from the exact one-particle results, indicating the importance of inter-particle interactions in the intermediate phase, in agreement with Ref.~\cite{pollmann_magnon}. 
This kind of picture is readily available in some extended Kitaev-like material $\alpha-RuCl_{3}$; here also authors claim multi-magnon processes are important in the high-field (field applied along $a$ and $b$-axis of the material)~\cite{Loidl, Valenti}.

\medskip

\noindent

\medskip

\noindent


\noindent{\it Magnon pair and density order parameters:}
In order to understand the processes that close the gap at the critical fields, we analyze the order parameter for forming a bound state of magnons or correspondingly, the boson bound-state order parameter
 $        \Delta^\gamma = (1/N)\sum_i \< a^{\dagger}_i a^{\dagger}_{i + {\delta\gamma}} \> 
$
and the boson density-density bond correlator
$ 
        N^\gamma = (1/N) \sum_i \<  n_i  n_{i + {\delta\gamma}} \>.
$

\begin{figure*}[t]
	\centering
	\subfloat{\label{fig:4a}}
    \subfloat{\label{fig:4b}}
    \subfloat{\label{fig:4c}}
    \subfloat{\label{fig:4d}}
    \subfloat{\label{fig:4e}}
    \includegraphics[trim={0cm 0cm 0 0},clip,width=0.98\textwidth]{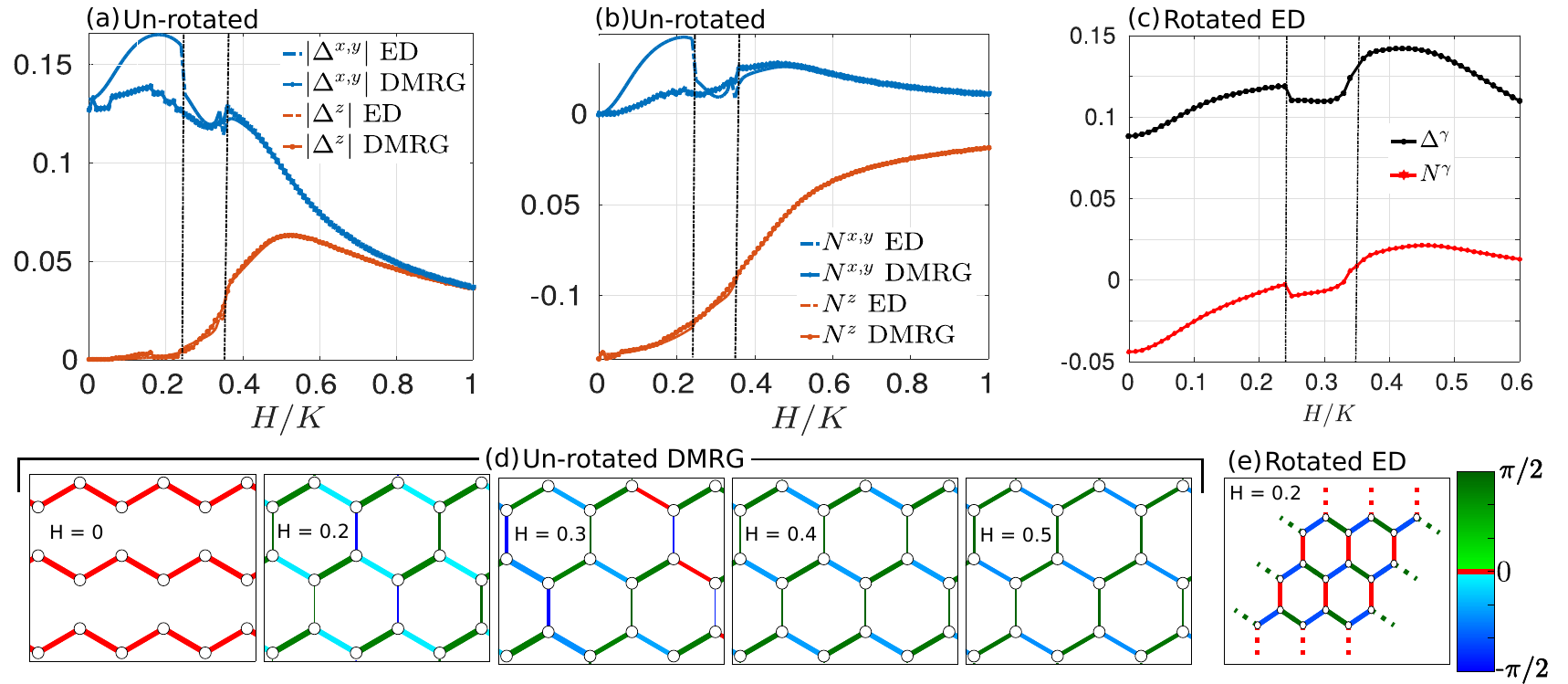}
    \vspace{-0.35cm}
	\caption{Results obtained using $16 \times 3$ unit cell ($96$ sites) DMRG in the spin representation and $3 \times 3$ unit cell ($18$ sites) ED for the hard core boson representation. Magnitude of spin-pairing $\Delta^{\gamma} = \Delta_0 e^{i\theta}$ on different bonds in the unrotated (x,y,z) coordinate system (panel a and d) and rotated ($\hat e_1,\hat e_2,\hat e_3$) coordinate system (panel c and e). The boson density-density bond correlations $\langle  n_i  n_j \rangle$ in the unrotated (panel b) and rotated (panel c) coordinates. ED and DMRG data agree qualitatively. Real space representation of pairing on different bonds on the honeycomb lattice for unrotated (panel d) and rotated (panel e) coordinates. Thickness of the line gives the magnitude $\Delta_0$ of the pairing and the phase $\theta$ is indicated by the color gradient.
	}
	\label{fig:4}
\end{figure*}

Fig.~\ref{fig:4a} shows that the pair order parameter is finite for the $x, y$-bonds since pairing operators appear explicitly 
in the Hamiltonian on the $x-, y-$ bonds since $\<  S_i^{+}  S_j^{+} \> \propto  \langle a_i^{\dagger}  a_j^{\dagger} \rangle$. On the other hand, two-boson pairing develops on the $z$-bond near the critical field $H_{c1}$ with a maximum value for $H \simeq 0.45$. The two-boson bound state is unfavourable in the polarized limit $(H \rightarrow \infty)$ as all correlations are suppressed in the product-state. 
Remarkably, two-particle pairing on all bonds is most favorable close to the phase boundary $H = H_{c2}$ of the gapless QSL and PPM phase. Fig.~\ref{fig:4d} is the real space representation of Fig.~\ref{fig:4a}, where the thickness of the lines represents the magnitude and the color represents the phase $\theta$. For $H = 0$, $\Delta^{x/y} \neq 0$ forms a zig-zag chain pattern with $\theta = 0$ phase. With increasing field, pairing develops on the $z$-bonds with $\theta > 0$ while the $x-$ and $y-$ bonds have the opposite phase structure for the pairing field.

At $H = 0$, we find $N^z \neq 0$ while $N^x = N^y = 0$ (fig.~\ref{fig:4b}). At the high field limit, $N^{\gamma}$ approaches zero. With increasing field, $N^z$ approaches zero monotonically while $N^{x/y}$ show non-monotonic behavior with $H$, with large density fluctuations near the critical field $H_{c2}$. 

We perform similar analysis of the pairing and boson density correlations using the `rotated' ($\{\hat{e}^1,\hat{e}^2,\hat{e}^3\}$) basis (Fig.~\ref{fig:4c}).
In the rotated basis $\Delta^\gamma$ and $N^\gamma$ on all bonds are equivalent as expected. 
There is sudden decrease in $\Delta^\gamma$ and $N^\gamma$ at $H_{c1}$ followed by a sharp increase at $H_{c2}$. The pair magnitude is largest beyond $H_{c2}$ followed by a continuous decrease into the partially polarized product state where all correlations are suppressed. The density correlations are real with no associated complex phases and are qualitatively similar to the local boson pair correlations. However, there is an overall change in the  sign associated with the $\<  n_i  n_j \>$ in the gapless phase. Fig.~\ref{fig:4e} shows the real-space representation of the boson pairing on different bonds for fixed values of $H$ analogous to fig.~\ref{fig:4d} for the unrotated case. For the rotated case, the phase on $x,y$-bonds remains equal and opposite and zero phase for the $z$-bond. Therefore, with increasing field only the amplitude of boson-pairing evolves and there is no change in the phase.\\
Is boson pair formation tied to the two spin-flip (SF) processes in the ground-state? We calculate the probability associated with different SF processes in the ground-state and the first excited state (Fig.~S2~\cite{suppl}). We discover that $2$ and $4$ SF processes are more likely and mix into the ground-state rather than $1$-SF. Further, the $1^{st}$ excited state in the PPM phase contains a high probability of even number of spin-flips for $H>0.5$. This clearly indicates that $2$-magnon excitations or pair-like excitations are key players in the phase transition near $H_{c2}$ into the gapless QSL phase. 

\bigskip

\noindent{\it Experimental Implications:} 
We predict that the low-energy Raman response for one-particle and two-particle excitations should show distinctive signatures in the different regimes of the phase diagram. One of our significant findings is that gap with decreasing magnetic field from the polarized regime closes by a pair-magnon process rather than by one magnon process. This result is confirmed from our calculations of the boson bound state on each bond, the spin-flip probability amplitude and one- and two- spin flip spectra. It would first be useful to establish that at fields much larger than $H_{c2}$, the two-spin gap is approximately twice the single-spin gap. Once this is established, it would be interesting to see if the gap scale for two-spin excitations dips below the one-spin excitation close to $H_{c2}$. 

\medskip

\noindent {\it Acknowledgements:} We acknowledge helpful discussions with Kyungmin Lee and Franz Utermohlen. Computations were performed using Unity cluster at The Ohio State University and the Ohio supercomputer. This work is supported by DOE grant DE-FG02-07ER46423.

\bibliography{references}

\clearpage
\pagebreak
\begin{widetext}
\clearpage
\begin{center}
\textbf{\large Supplementary Information For \\
``Two-Magnon Bound States in the Kitaev Model in a $[111]$-Field''}
\end{center}
\end{widetext}

\setcounter{equation}{0}
\setcounter{figure}{0}
\setcounter{table}{0}
\setcounter{page}{1}
\makeatletter
\renewcommand{\theequation}{S\arabic{equation}}
\renewcommand{\thefigure}{S\arabic{figure}}

\noindent
{\bf S.1 Hard-core boson transformation and axis rotation} --- 

Kitaev $S = 1/2$ Hamiltonian in the presence of a magnetic field along [111] is defined as
\begin{equation} \label{Seq:Kitaev_suppl}
H_{K} =  \sum_{\gamma = x, y, z} K^{\gamma} \sum_{\< ij \>} S_i^\gamma S_j^\gamma - \mathbf{H} \cdot  \sum_{i, \gamma} \mathbf{S_i^{\gamma}},
\end{equation}
where operator $S^{\gamma} = 1/2 \sigma^{\gamma}$, where $\sigma$ are Pauli matrices. The $S = 1/2$ spin operators can be mapped exactly to hard core bosons (HCB) through this transformation:
\begin{equation}
\begin{split}
 S_{i}^{+} &=  a_{i}^\dagger, \ \ \ \ \ \ \ \ \ \ \ \ \ \  S_{i}^{-} =  a_{i}, \\
 S_{i}^{z} &= n_{i} - \frac{1}{2} \ \ \ \ \ \ \ \  n_{i} = a^\dagger_{i} a_{i},
\end{split}
\end{equation}
that satisfy commutation relations:
\begin{equation}
\begin{split}
[ S_i^+,  S_j^{-}] &= 2\delta_{ij}  S_i^z,  \\
[ S_i^{\pm},  S_j^{z}] &= \mp\delta_{ij}  S_i^{\pm} \\
[ a_i,  a_j^{\dagger}] &= \delta_{ij}(1 - 2  a_i^{\dagger}  a_i)
\end{split}
\end{equation}
\noindent
with an on-site exclusion principle $(S_i^{\pm})^2 = 0$; 
$ S_{i}^{\pm} =  S_i^{x} \pm i S_i^{y}$.
The $a_{i}^\dagger$ ($a_{i}$) are boson creation (annihilation) operators on a site $i$.
The one-spin- flip operator $S_i^{+}$ is equivalent to creating a boson, and 
the exclusion principle leads to a constraint $n_i =  0$ or $1$, hence a hard-core boson constraint. In the HCB transformed basis, the up spin is identified as singly occupied boson and a down spin is identified as an empty site. This analogy between the spin variables and the HCB variables was first explored in the context of helium~\cite{HCB} and later in the context of the Heisenberg model~\cite{Nandini_Ceperley}. 

The above Hamiltonian~(\ref{Seq:Kitaev_suppl}) can be expressed in terms of HCB in the $\{ \hat{e_x}, \hat{e_y},  \hat{e_z} \}$ (`unrotated') bond-directional basis as

\begin{equation}
\begin{split}
H_{K} &= \frac{K^x}{4}\sum_{\langle ij \rangle_x}(a_i^\dagger a_{j} +  a_i^\dagger a_{j}^\dagger + h.c.)\\
&+ \frac{K^y}{4}\sum_{\langle ij \rangle_y}( a_i^\dagger  a_{j} -  a_i^\dagger  a_{j}^\dagger +h.c.)\\
&+ K^z\sum_{\langle ij \rangle_z}( n_i  n_{j} -  n_i S -  n_{j} S)\\
&-H\sum_{i}[ \frac{1}{2}(1+i) a_i^\dagger +  \frac{1}{2}(1-i) a_i +  n_i]\\
&+ \sum_{\langle ij \rangle_z}K^z S^2 - \sum_{\langle i\rangle}HS
\end{split}
\end{equation}

\noindent
$\langle i j\rangle$ are nearest neighbours (provided $i <j $) corresponding to different bonds $\gamma = {x, y, z}$. $a_i^\dagger a_{i+\gamma}$ describe nearest neighbour hopping, $a_i^\dagger a_{i + \gamma}^\dagger$ denotes pairing defined on a bond $\gamma$ and $S=1/2$. Four boson term $n_i n_j$ appears on the $z$-bond only.
In addition to HCB transformation, we also perform a coordinate transformation 

\begin{equation}\label{Seq:basis_transform}
\begin{split}
    \mathbf{S^{e}} &= R \ \mathbf{S^{\gamma}} \\
 \begin{bmatrix} 
    S^{1} \\ 
    S^{2} \\
    S^{3}
 \end{bmatrix}
 &=
  \begin{bmatrix}
   -\frac{1}{\sqrt{6}} & -\frac{1}{\sqrt{6}} & \frac{2}{\sqrt{6}}\\
   \frac{1}{\sqrt{2}} & -\frac{1}{\sqrt{2}} & 0 \\
   \frac{1}{\sqrt{3}} & \frac{1}{\sqrt{3}} & \frac{1}{\sqrt{3}} 
   \end{bmatrix}
    \begin{bmatrix} 
    S_{x} \\ 
    S_{y} \\
    S_{z}
 \end{bmatrix}. 
\end{split}
\end{equation}

such that the $z$-spin projection is aligned along the direction of the magnetic field: $ \frac{1}{\sqrt{3}} \ (\hat{e}^x+\hat{e}^y+\hat{e}^z)$. In the `rotated' basis, the up spin configuration refers to spin aligned along the $[111]$-direction of the original $\{\hat{e}^x,\hat{e}^y,\hat{e}^z\}$ basis. This coordinate rotation allows us to analyze the spin-flip processes in the ground-state.

The Hamiltonian in the new `rotated basis' $\{ \hat{e^1},\hat{e^2}, \hat{e^3}\}$ with the HCB transformation is 

\begin{equation} \label{Seq:rotatedHCB}
\begin{split}
H_{K'} &=  \frac{K^{x}}{6}\sum_{ {\langle ij \rangle}_{x}}(\lambda  a_i^\dagger  a_{j}^\dagger +  a_i^\dagger  a_{j} + h.c.)\\
&-\sqrt{2}{K^x} \sum_{ {\langle ij \rangle}_{x}}(\lambda^*  a_i^{\dagger}  a_i  a_j^{\dagger} + \lambda  a_i^\dagger  a_i  a_j +h.c.)\\
&+ \frac{K^{y}}{6}\sum_{ {\langle ij \rangle}_{y}}(\lambda^*  a_i^\dagger  a_j^\dagger +  a_i^\dagger  a_j+h.c.)\\
&-\sqrt{2}{K^y} \sum_{ {\langle ij \rangle}_{y}}(\lambda  a_i^{\dagger}  a_i  a_j^{\dagger} + \lambda^*  a_i^\dagger  a_i  a_j +h.c.)\\
&+ \frac{K^{z}}{6}\sum_{ {\langle ij \rangle}_{z}}( a_i^\dagger  a_j^\dagger +  a_i^\dagger  a_j+h.c.)\\
&-\sqrt{2}{K^z}\sum_{ {\langle ij \rangle}_{z}}( a_i^{\dagger}  a_i  a_j^{\dagger} +  a_i^\dagger  a_i  a_j +h.c.)\\
& +2\sum_{ {\langle ij \rangle}_{\gamma}}{\frac {K^{\gamma}} {6}}( a_i^\dagger  a_i  a_j^\dagger  a_j)-3S\sum_{{i}}( a_i^\dagger  a_i- S)\\
&-\sqrt{3}H \sum_{{i}}(S/2+  a_i^\dagger  a_i), 
\end{split}
\end{equation}

\noindent
where $\lambda = -\frac{1}{2}(1+i\sqrt{3})$ with $\lambda\lambda^* = 1$ and $\lambda + \lambda^* +1 = 0$.
In the `rotated' basis all bonds becomes equivalent as it comprises similar terms on all bonds and in addition three boson terms appear on all bonds and these terms also contribute to the dynamics of the system.
\vspace{0.40cm}

{\bf S.2 Benchmarking Results with HCB Representation} --- 

We analyze the rotated HCB Kitaev model (\ref{Seq:rotatedHCB}) using ED and Lanczos~\cite{Lanczos}, while we directly simulate the Kitaev model in the un-rotated spin representation (\ref{Seq:Kitaev_suppl}) using DMRG. We use $N = 8, 12, 16, 18$ and $96$ sites honeycomb geometry on a torus for ED/Lanczos and a cylinder for DMRG. We begin by analyzing the ground-state energy, magnetization $M$, and spin susceptibility $\chi$ defined as, 
\begin{equation}
    \begin{split}
        M &= \frac{1}{N\sqrt{3}} \sum_{i} (\<S^x_i\> + \<S^y_i\> + \<S^z_i\>) \\
        &= \frac{1}{N}  \sum_{i}  \<S^3_i\> \\
        \chi &= \frac{\partial M}{\partial H}
    \end{split}
\end{equation}
to benchmark our results with previous findings~\cite{Nirav, Trebst, David, Sheng, Pollmann}. Note that $M$ is defined in the original $\{\hat{e}^x,\hat{e}^y,\hat{e}^z\}$ bond basis and can also be obtained in the $\{\hat{e}^1,\hat{e}^2,\hat{e}^3\}$ basis using Equation~\ref{Seq:basis_transform}. 

\begin{figure}
	\centering
	\includegraphics[trim={0 0.0cm 0 0},clip, width=0.43\textwidth]{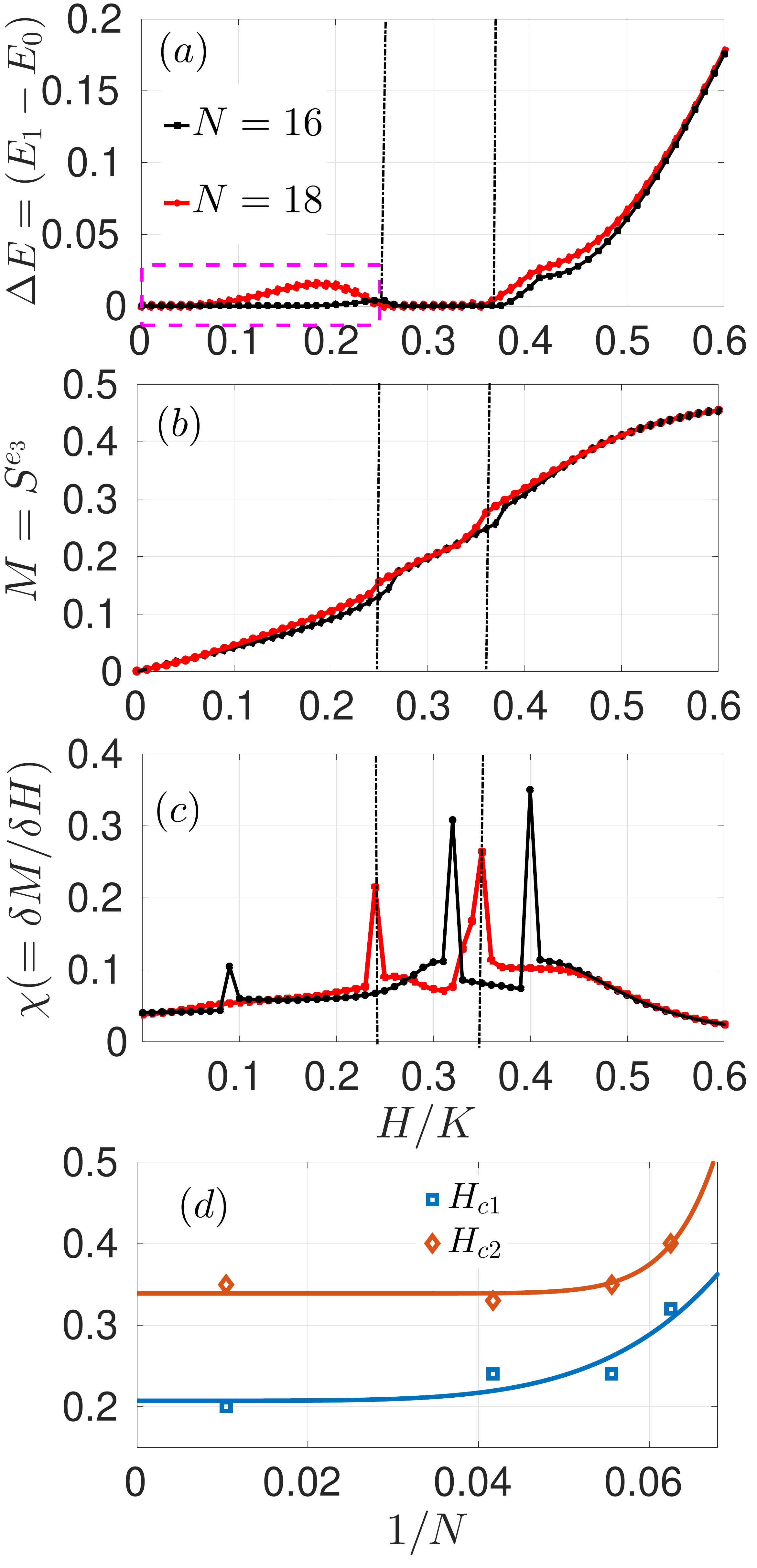}
	\subfloat{\label{fig:s1a}}
    \subfloat{\label{fig:s1b}}
    \subfloat{\label{fig:s1c}}
    \subfloat{\label{fig:s1d}}
	\caption{
	(a) Energy gap ($\Delta E = E_1 -E_0$), (b) magnetization $M$ and (c) spin susceptibility $\chi$ as a function of the magnetic field for $16$ and $18$ sites. Magnetization $M$ is defined along the rotated $\mathbf{\hat{e}^3}$ direction. The two peak positions in $\chi$ are marked by two black dashed lines indicating two critical fields $H_{c1} =\simeq 0.24$ and $H_{c2} = \simeq 0.35$. 
    (d) Finite-size scaling of the critical fields $H_{c1}$ and $H_{c2}$ as a function of $1/N$, where, $N$ is the number of sites ($N = 16, 18, 24, 96$) where $N = 24$ and $96$ data are obtained from DMRG~\cite{Nirav}. The extrapolated values of $H_{c1}$ and $H_{c2}$ are 0.208 ($\pm 0.03$) and 0.340 ($\pm 0.01$) respectively in the $N \rightarrow \infty$ limit.}
	\label{fig:s1}
\end{figure}

The energy gap ($\Delta E = E_1 - E_0$) as a function of the magnetic field is presented in Fig.~\ref{fig:s1a}. 
With decreasing field, the energy gap closes at a critical field strength $H_{c2}$ and the spectrum remains gapless upto $H_{c1}$. 
For $H \leq H_{c1}$, a gap reopens and vanishes again at $H = 0$. 

The magnetization in the presence of $[111]$-field would tend to align the spins along the $\hat{e}^3$ in the rotated basis, therefore we find that $S^{3} = 0.5$ and $S^{1} = S^{2} = 0$ for large fields. The two-step structure of $M$ in Fig.~\ref{fig:s1b} and the corresponding two-peak structure of $\chi$ in Fig.~\ref{fig:s1c} indicate two-phase transitions, in agreement with our previous works~\cite{Sheng,Nirav,David}.
A finite-size scaling analysis in Fig.~\ref{fig:s1d} using a combined ED, Lanczos and DMRG data yields extrapolated values of $H_{c1} = 0.208\pm 0.03$ and $H_{c2} = 0.340\pm 0.01$.

\begin{figure*}[t]
	\centering
		\includegraphics[width=0.85\textwidth]{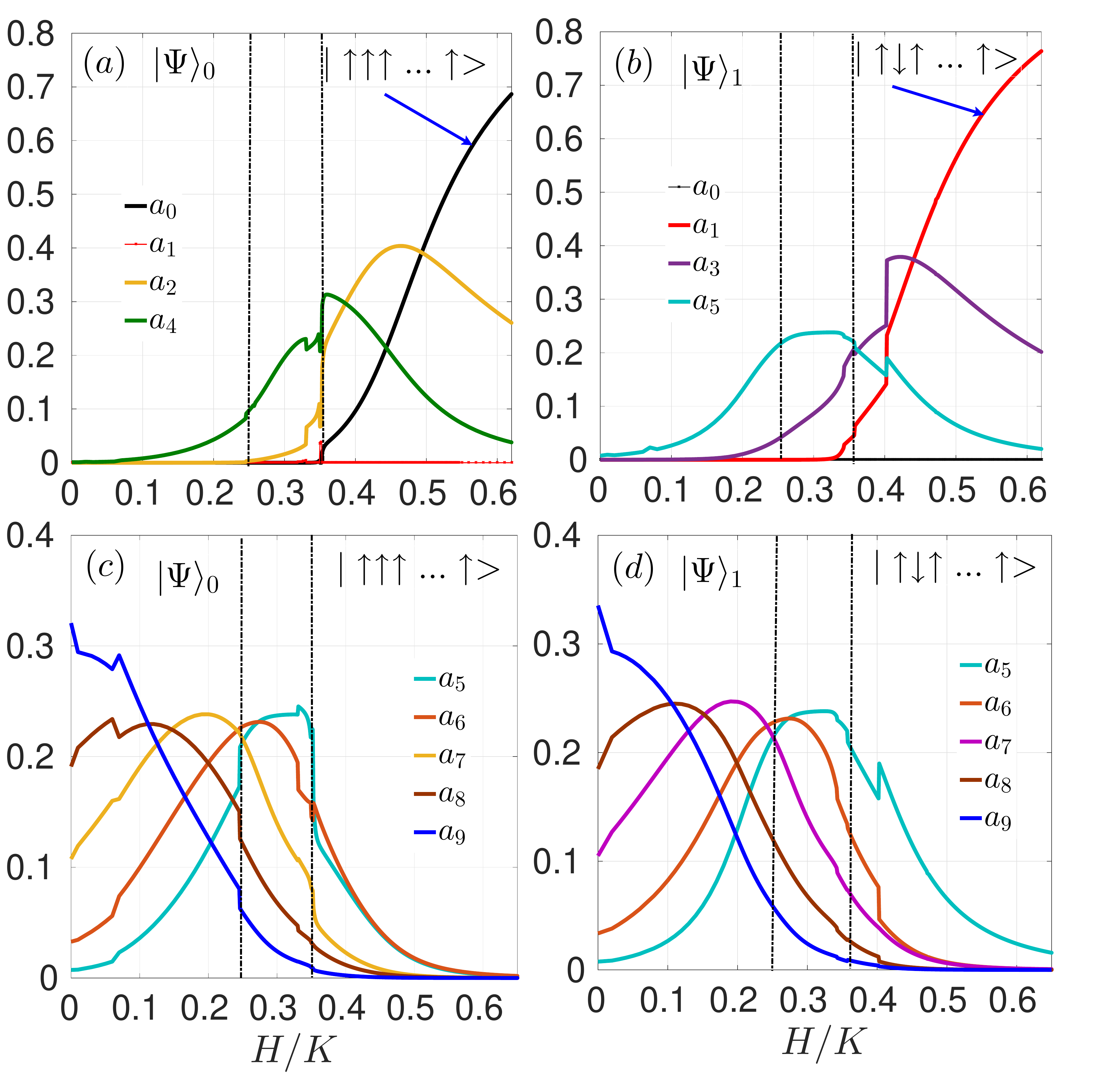}
	\subfloat{\label{fig:s2a}}
    \subfloat{\label{fig:s2b}}
    \subfloat{\label{fig:s2c}}
    \subfloat{\label{fig:s2d}}
	\caption{
	Dominant spin-flip processes that contribute to the closing of the magnon gap from the ground state and the first excited state wave functions respectively. (a) and (b) show the results from the high-field end; (c) and (d) show the results from the low-field Kitaev spin liquid end.}
	\label{fig:s2}
\end{figure*}
\vspace{0.40cm}

{\bf S.3 Spin-flip Probabilities} ---

We address the following questions:

\noindent (1) What are the dominant processes involved in closing the spin gap with decreasing $H$? 

\noindent (2) Is boson bound-state formation tied to the two spin-flip processes in the ground-state? 

To this end, we show probability associated with `$n$' spin-flips in the ground-state $|\Psi_0\>$ and the first excited state $|\Psi_1\>$ (Fig.~\ref{fig:s2}). 
We define this probability as 
\begin{equation}
   a_{n} = \sum_{b \in n} |c_{b}|^2 \ \ \ \ \ 
\end{equation}
where $N$ is the number of sites, $n = N - \<b|S^z_{total} |b\>$ is the number of spin flips relative to the fully polarized phase, and $b$ labels a particular configuration of the $N$ spins with the constraint that the configuration has $n$ spin flips. For example, the ground-state is $\ket{\uparrow \uparrow \uparrow .... \uparrow}$ in a polarized phase at $H \rightarrow \infty$. With zero spin-flip processes, this is the only candidate because only configuration $\ket b = \ket{\uparrow \uparrow \uparrow.... \uparrow }$ contributes to the ground-state. This in turn implies that the probability associated with $n \neq 0$ spin-flips are zero ($a_{n \neq 0} = 0$) as shown in Fig.~\ref{fig:s2a}. 

\noindent
\underline{High-field phase:} The first excited state in the PPM phase corresponds to creating a magnon or a single spin-flip. Fig.~\ref{fig:s2b} shows that indeed the $1^{st}$ excited state for $H>0.5$ has a large probability associated with $a_1$, i.e., a single spin-down in a sea of spin-up ($1$ spin-flip). 
Naively, one would expect that upon decreasing $H$ towards $H_{c2}$, the ground-state probability of $1$ spin-flip process would dominate. Remarkably, we discover that two and four spin-flip ($a_2$ and $a_4$) processes become more likely in the ground-state rather than $a_1$ (Fig.~\ref{fig:s2a}). This clearly indicates that two-magnon excitations or pair-like excitations are the key players in the phase transition near $H_{c2}$. Similar analysis of the $|\Psi_1\>$ shows that the $a_3$ process dominates near $H_{c2}$, i.e., 2-flips with respect to the high-field $|\Psi_1\> = |\uparrow \downarrow \uparrow ... \uparrow \>$. This analysis shows the dominance of even spin-flips, with respect to the high field state, as the phase transition to the gapless QSL phase is approached. 

\noindent
\underline{Low-field phases:} The zero field KSL phase preserves time-reversal symmetry and hence we expect an equal number of spin-up and spin-down configurations mixing in the ground-state. This is in agreement with the large probability associated with $a_9$, i.e., $N/2=9$ for the $N=18$ spins considered here in the up-state along the ${e}^3$ direction as shown in Fig.~\ref{fig:s2c}. 
The linear superposition of configurations with equal number of up and down spins is also consistent with a spin-disordered QSL ground state and also for the excited state $|\Psi_1\>$ (Fig.~\ref{fig:s2d}). Upon increasing $H$, the system develops a finite magnetization as spin-up configurations become more favorable relative to spin-down. Therefore, we find an overall shift in the peaks of $a_n$ with increasing $H$, i.e., the higher order spin-flip processes are less likely with increasing $H$ (Fig.~\ref{fig:s2c} and Fig.~\ref{fig:s2d}). \\

\vspace{0.4cm}
\noindent
{\bf S.4 Linear Spin-Wave Theory } ---
The high-field polarized phase hosts ``magnons" as topological excitations~\cite{Joshi, magnons_kitaev}. Spin waves in magnetically ordered systems are analog of lattice waves in solid systems, where a quantized spin wave is called a ``magnon". These are best studied within the spin-wave theory by representing the spin operators in terms of auxiliary bosons via the Holstein-Primakoff transformation. We expand about this fully polarized state using the following transformation and commutation relations: 

\begin{equation}\label{Seq:Holstein-Primakoff}
\begin{split}
 S_j^z = S-  n_j \ \ \ \ \ \ \left[ a_i,  a_j^\dagger \right ] &= \delta_{ij} \\
 S_j^+ = \sqrt{(2S-  n_j)} a_j \ \ \ \ \ \ \left[ a_i,  a_j\right] &= 0 \\
 S_j^- =   a_j^\dagger \sqrt{(2S-  n_j)} \ \ \ \ \ \ \left[ a_i^\dagger,  a_j^\dagger\right] &= 0. 
\end{split}
\end{equation}

We restrict ourselves to linear spin-wave theory wherein we keep only the bilinear terms (systematic expansion parameter upto $1/S$) in the bosonic Hamiltonian. Such an approximation is controlled for large $S$.
Note that the transformation (\ref{Seq:Holstein-Primakoff}) can be viewed as an $1/S$ expansion in the magnon density $\rho =  \frac{\langle b_i^\dagger b_i\rangle}{(2S)}$, controlled in the limit, $\rho < 1$.

\begin{equation}\label{eq:Holstein-Primakoff2}
\begin{split}
 S_i^z &= S -  a_i^\dagger  a_i\\
 S_i^+ &= \sqrt{2S} \sqrt{1 - \frac{ a_i^\dagger  a_i}{2S}}  a_i\simeq \sqrt{2S} ({1 - \frac{ a_i^\dagger  a_i}{4S}})  a_i\\
 S_i^- &= \sqrt{2S}  a_i^\dagger \sqrt{1 - \frac{ a_i^\dagger  a_i}{2S}}\simeq \sqrt{2S}  a_i^\dagger ({1 - \frac{ a_i^\dagger  a_i}{4S}}).
\end{split}
\end{equation}

\noindent
Keeping only the quadratic terms, the resultant Kitaev Hamiltonian after the Holstein-Primakoff transformation in the rotated basis is given by,

\begin{equation}\label{eq:LSWT hamiltonian}
\begin{split}
H_{SW} &=  \sum_{i} \big[KS( a_i^\dagger  a_{i+\delta_x} +  a_i^\dagger  a_{i+\delta_y}
+  a_i^\dagger  a_{i+\delta_z} \\
&+ \lambda^*  a_i a_{i+\delta_x} + \lambda  a_i a_{i+\delta_y}
+  a_i a_{i+\delta_z}+ \text{h.c.}) \\
&+ S^2-\sqrt{3}H \sum_{{i}}(S/2+  a_i^\dagger  a_i) \\
&-3S\sum_{{i}}( a_i^\dagger a_i- S) \big] \\
\end{split}
\end{equation}

\noindent
here, $\lambda = -\frac{1}{2}(1+i\sqrt{3})$ with $\lambda\lambda^* = 1$ and $\lambda + \lambda^* +1 = 0$. The vectors to the nearest neighbor sites are defined as $\mathbf{\delta_x} = \frac{a}{2}(1, \frac{1}{{\sqrt{3}}})$, $\mathbf{\delta_y} = \frac{a}{2}(1,-\frac{1}{{\sqrt{3}}})$, $\mathbf{\delta_z} = a(0, \frac{1}{{\sqrt{3}}})$
in $\{\hat{e}^1,\hat{e}^2,\hat{e}^3\}$ coordinates with $a=1$. The spin-wave 
Hamiltonian is obtained by keeping only the quadratic term and using Fourier transform of the boson operators

\begin{equation}\label{eq:SWT_hamiltonian}
\begin{split}
a_k &= \frac{1}{\sqrt{N}}\sum_{i} a_i e^{i {\mathbf k}.{\mathbf r_i}}, \\
\Psi_k &= (a_\mathbf{k}, b_\mathbf{k}, a_{-\mathbf{k}}^\dagger, b_{-\mathbf{k}}^\dagger), \\
H_{SW} &= \frac{1}{2} \sum_{\mathbf{k}}\Psi_\mathbf{k}^\dagger H_{SW} (\mathbf{k}) \Psi_\mathbf{k}, \\
H_{SW}(\mathbf{k}) &= 
\begin{bmatrix}
M(\mathbf{k}) & N(\mathbf{k})\\
M(\mathbf{k})^\dagger & N(-\mathbf{k})^T\\
\end{bmatrix},
\end{split}
\end{equation}

\FloatBarrier
\begin{figure}[b]
\centering
\includegraphics[trim={0cm 0cm 0cm 0cm},width=1\linewidth]{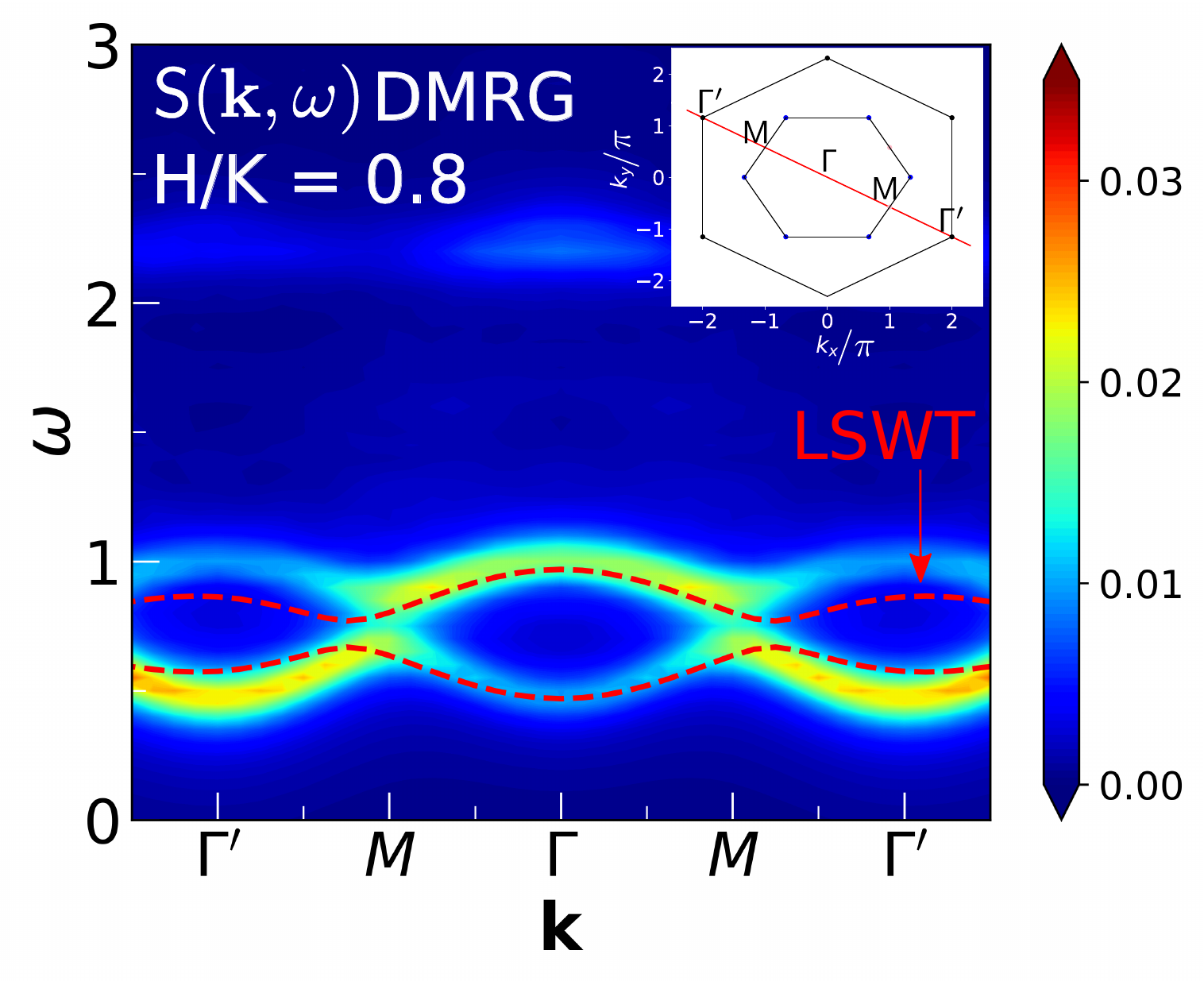}
\vspace{-0.7cm}
\caption{One-magnon spectral function $S(\mathbf{k},\omega)$ calculated using DMRG (countour) and linear spin-wave theory (red dashed line). Presented results are along a cut of the Broullion zone, as shown by the inset where corresponding high symmetry points are labeled. The DMRG results are obtained using a $16 \times 3$ unit cell with the correction-vector formulation in
Krylov space~\cite{Krylov1,Krylov2}. LSWT results are for infinite 2D system. All results are obtained using $H/K=0.8$.}
\label{fig:s3}
\end{figure}

\noindent
The $M(k)$ and $N(k)$ are defined as 
\begin{equation}
\begin{split}
M(k) &= 
\begin{bmatrix}
-KS+H & A \\
 A^* & -KS+H \\
\end{bmatrix}, \\
N(k) &= 
\begin{bmatrix}
0 & B\\
 C & 0\\
\end{bmatrix},
\end{split}
\end{equation}
where $A = \frac{KS}{3}(e^{i {\mathbf k} \cdot \mathbf{\delta_x}} + e^{i {\mathbf k} \cdot \delta_y} +e^{i{\mathbf k} \cdot \delta_z})$,  
$B = \frac{KS}{3}(\lambda ^* e^{i{\mathbf k} \cdot \mathbf{\delta_x}} + \lambda e^{i{\mathbf k} \cdot \mathbf{\delta_y}} + e^{i{\mathbf k} \cdot \mathbf{\delta_z}})$ and 
$C = \frac{KS}{3}(\lambda ^* e^{-i{\mathbf k} \cdot \mathbf{\delta_x}} + \lambda e^{-i{\mathbf k} \cdot \mathbf{\delta_y}} +e^{-i{\mathbf k} \cdot \mathbf{\delta_z}})$. 

\vspace{0.20cm}

Fig.~\ref{fig:s3} shows our results for the one-magnon spectral function calculated using DMRG for a certain cut along the Brillouin zone at $H/K = 0.80$ along with the linear spin wave theory results. We find good agreement at large fields, as expected.


\end{document}